%% file: ASKAP_reps.tex
\documentclass[fleqn,usenatbib,useAMS]{mnras}

\usepackage{newtxtext,newtxmath}

\usepackage[T1]{fontenc}
\usepackage{ae,aecompl}

\newcommand{\pccc}{\,pc\,cm$^{-3}$}

\usepackage{graphicx}	
\usepackage{amsmath}	
\usepackage{amssymb}	

\title[Which bright fast radio bursts repeat?]{Which bright fast radio bursts repeat?}

\author[C.W.~James et al.]{
C.W.~James,${^{1}}$\thanks{E-mail: clancy.james@curtin.edu.au}
S. Os\l{}owski,$^{2}$\thanks{E-mail: stefanoslowski@swin.edu.au} 
C.~Flynn,${^2}$
P.~Kumar,${^2}$ 
K.~Bannister,${^3}$
S.~Bhandari,${^3}$ \newauthor
W.~Farah,${^2}$
M.~Kerr,${^4}$
D.R.~Lorimer,${^{5,6}}$ 
J.-P.~Macquart,${^{1}}$ 
C.~Ng,${^7}$
C.~Phillips,${^3}$ \newauthor 
D.C.~Price,${^{2,8}}$ 
H.~Qiu,${^{3,9}}$ 
R.M.~Shannon${^2}$ and
R.~Spiewak${^{2,10}}$ \\
\input{address}}

\date{Accepted XXX. Received YYY; in original form ZZZ}

\pubyear{2019}

\begin{document}
\newcommand{\repeater}{\mbox{FRB~121102}}

\label{firstpage}
\pagerange{\pageref{firstpage}--\pageref{lastpage}}
\maketitle

\begin{abstract}
A handful of fast radio bursts (FRBs) are now known to repeat. However, the question remains --- do they all? We report on an extensive observational campaign with the Australian Square Kilometre Array Pathfinder (ASKAP), Parkes, and Robert C.\ Byrd Green Bank Telescope, searching for repeat bursts from FRBs detected by the Commensal Real-time ASKAP Fast Transients survey. In 383.2\,hr of follow-up observations covering 27 FRBs initially detected as single bursts, only two repeat bursts from a single FRB, FRB~171019, were detected, which have been previously reported by Kumar et al. We use simulations of repeating FRBs that allow for clustering in burst arrival times to calculate new estimates for the repetition rate of FRB~171019, finding only slight evidence for incompatibility with the properties of FRB~121102.
Our lack of repeat bursts from the remaining FRBs set limits on the model of all bursts being attributable to repeating FRBs. Assuming a reasonable range of repetition behaviour, at most 60\% (90\%\,C.L.) of these FRBs having an intrinsic burst distribution similar to FRB~121102.
This result is shown to be robust against different assumptions on the nature of repeating FRB behaviour, and indicates that if indeed all FRBs repeat, the majority must do so very rarely.
\end{abstract}




\section{Introduction}

Fast Radio Bursts (FRBs) are mysterious, bright bursts of radiation at radio wavelengths, discovered serendipitously just over a decade ago by 
\citet{Lorimer2007} at the Parkes telescope. FRBs have durations of a few hundred microseconds to tens of milliseconds, and dispersion measures (DMs) which can exceed the contribution due to the Interstellar Medium (ISM) of the Milky Way by more than an order of magnitude. Their high DMs are dominated by traversal of the pulse through the Intergalactic Medium (IGM) over cosmological distances \citep{Shannonetal2018}. At the time of writing, there are nearly 100 FRBs listed in the FRB catalog at {\tt frbcat.org} \citep{Petroffetal2016}. The progenitors of FRBs are currently unknown, with almost as many theories for their origins as there are observed FRBs (see \cite{Platts2018} and {\tt frbtheorycat.org} for an extensive review). 

To date, FRB observations from seven different facilities have been published. Chronologically, these are the Parkes radio telescope \citep{Lorimer2007}, the Robert C.\ Byrd Green Bank Telescope \citep[GBT; ][]{GBT}, the Arecibo Observatory \citep{Spitler2014}, the upgraded Molonglo synthesis telescope \citep[UTMOST; ][]{UTMOSTFRBs}, the Australian Square Kilometre Array Pathfinder \citep[ASKAP; ][]{Bannisteretal2017}, the Canadian Hydrogen Intensity Mapping Experiment \citep[CHIME; ][]{CHIME2019a}, and the Deep Synoptic Array ten-antenna prototype \citep[DSA-10; ][]{2019Natur.572..352R}.
FRBs have been seen at radio frequencies only, from 400\,MHz to 8\,GHz, despite extensive follow-up work at all other wavelengths \citep[e.g.][]{Bhandarietal2018}.

No additional bursts were discovered at the location of any FRB until 9 repetitions from FRB~121102 were discovered at the 305-m Arecibo telescope \citep{Spitler2016}, in a source itself discovered earlier by \citet{Spitler2014}. Since then, \citet{CHIME2019a,CHIME2019b,2020ApJ...891L...6F} have discovered 18 new repeating sources.
The inferred all-sky rate of bursts observed by CHIME is too high to be consistent with models predicting once-off bursts \citep{Ravi2019}. Most recently, in GBT follow-up observations, one of the ASKAP bursts, FRB~171019, has been observed to repeat \citep{Kumar2019}.

Despite these recent discoveries, \repeater, having been localised to its host galaxy at $z=0.1927$ \citep{Tendulkar2017,Chatterjee2017}, remains by-far the best-studied FRB. The degree to which the properties of \repeater\ relate to the other repeaters is currently unclear. At a more fundamental level, the question ``which FRBs repeat?'' remains unanswered, with an alternative hypothesis allowing for two or more populations.

Here, we test this hypothesis on the bright FRB population through deep follow-up observations for FRBs detected by The Commensal Real-time ASKAP Fast Transients (CRAFT) Survey.
In Section~\ref{sec:observations}, we describe the full set of observations from this follow-up program. Using a model for a repeating FRB developed in Section~\ref{sec:model}, we place limits on repetition rates --- allowing for non-Poissonian burst-wait-time distributions --- in Section~\ref{sec:limits}. Importantly, by only analysing the probability of detecting multiple bursts from already identified FRBs, we eliminate the bias inherent in the initial detection.
FRB~171019 is analysed similarly in Section~\ref{sec:FRB171019}, where we derive both lower and upper bounds on its repetition rate. We discuss our results in the wider context of models of repeating FRBs in Section~\ref{sec:discussion}.

\section{CRAFT observation program}
\label{sec:observations}

CRAFT is a very wide-area FRB survey using the Australian Square Kilometre Array Pathfinder (ASKAP) telescope \citep{2010PASA...27..272M}. At the time of writing, 28 FRBs have been reported \citep{Bannisteretal2017,Shannonetal2018,Macquart2019a,Qiu2019,Bhandarietal2019,Agarwal2019,2019Sci...365..565B,2019Sci...365.0073P}. The automated FRB search scheduling, and very wide search area on the sky, means that the locations of most detected FRBs have been observed extensively. This alone can be used to constrain the repetition rate \citep{Bhandarietal2019,James2019b}. No repeat bursts have been found for any previously reported FRBs in ASKAP data, which is inconsistent with all FRBs sources being similar to FRB~121102 \citep{James2019b}.

Compared to other FRB searches, ASKAP/CRAFT observations have a relatively high detection threshold --- 26\,Jy\,ms in Fly's Eye mode \citep{Shannonetal2018,2019PASA...36....9J}, and no lower than 4.3\,Jy\,ms in incoherent sum mode \citep{2019Sci...365..565B}. This is compensated for by having a high field of view (FOV), making it sensitive to the rarest, and (likely intrinsically) brightest, bursts. We have therefore been pursuing an extensive follow-up program of CRAFT FRBs with Parkes and the GBT. These are both more sensitive telescopes, the limited FOVs of which are offset by CRAFT FRBs being localised to a few arcmin. They are also capable of observing a similar frequency range to that over which CRAFT FRBs have been discovered (0.8--1.4\,GHz).

This CRAFT follow-up program has recently discovered two repeat bursts from one of the bright ASKAP/CRAFT FRBs, FRB~171019 \citep{Kumar2019}, with the CHIME collaboration also detecting a repeat burst \citep{CHIME2019d}.  Additionally, one new burst, FRB~180318, has been discovered, which is unrelated to any of the previously observed bursts. The analysis of this FRB is ongoing, and will be reported elsewhere. However, no repeat bursts have been detected from any of the other CRAFT FRBs.

\subsection{Follow-up observations}

\begin{table*}
    \centering
    \begin{tabular}{c c c c c c c }
        \hline
    Telescope    & Receiver/      & $\bar{\nu}$   & $\Delta \nu$  & $\delta t$    & $\delta \nu$  & $F_{\rm th}$ \\
                & Mode   & [MHz]         & [MHz]         &   [ms]          & [MHz]           & [Jy\,ms] \\
       \hline
       ASKAP    &  FE  & 1315       & 336         & 1.2565        & 1             & 21.9 \\
                  & ICS         & 864--1320 & 336 & 0.864--1.728              & 1             & $21.9 N_{\rm ant}^{-0.5}$ \\
       \hline
       Parkes   & MB  & 1382         & 337.1            & 0.064         & 0.39          & 0.5 \\
       \hline
       GBT      & 820 MHz   &  820         &   200            & 0.08192       & 0.0977     & 0.12 \\
             & L-band   &  1500         &   800            & 0.08192       & 0.0977     & 0.058 \\
       \hline
    \end{tabular} 
    \caption{Observational properties of follow-up observations, for ASKAP Fly's Eye (FE) and incoherent sum (ICS) modes \citep{Shannonetal2018,2019Sci...365..565B}; the Parkes multibeam (MB) receiver \citep{SUPERB1}; and the Greenbank Telescope's (GBT's) 820\,MHz primary focus and L-band receivers \citep{Kumar2019}. From left to right: the telescope and receiver names, the central frequency $\bar{\nu}$ and total bandwidth $\Delta \nu$, time- and frequency- resolutions $\delta t$ and $\delta \nu$, and nominal sensitivity to a 1\,ms duration burst.}
    \label{tab:telescope_properties}
\end{table*}

The data reported here cover ASKAP, Parkes, and GBT observations up to June $26^{\rm th}$ 2019, targeting the first 27 FRBs reported by the CRAFT collaboration up to and including FRB~180924. Observations at Parkes were recorded using the multibeam receiver \citep{multibeam}, which nearly overlaps with the CRAFT observations in terms of frequency coverage. The observations at GBT were performed mostly used the 800 MHz receiver, with a few observations performed at 1500 MHz. The Parkes observations were analysed in real-time using the standard transient pipeline based on \textsc{Heimdall} \citep{Barsdell2012} used by other surveys, most recently by \citet{Oslowski2019}. GBT data were recorded for offline processing and searched for repeated bursts using the pipeline described in detail by \citet{Kumar2019}. Table~\ref{tab:telescope_properties} summarises the relevant properties of the telescopes and Table~\ref{tab:observations_sum} shows the total effective amount of time observed per source with the different follow-up facilities included in this analysis.

\begin{table}
	\centering
	\begin{tabular}{cccccc}
		\hline
		& \multicolumn{2}{c}{ASKAP} & Parkes & \multicolumn{2}{c}{GBT} \\ 
		FRB & FE & ICS & MB & 820 & L \\
		\hline
170107 & 883.5 & 7.3 & 30.0 & 10.9 & 4.3 \\
170416 & 482.6 & 1.3 & 15.8 & 0.0 & 0.0 \\
170428 & 912.9 & 1.5 & 12.2 & 3.5 & 1.3 \\
170707 & 343.4 & 0.8 & 1.8 & 0.0 & 0.0 \\
170712 & 205.7 & 2.1 & 4.6 & 0.0 & 0.0 \\
170906 & 1148.4 & 3.6 & 4.1 & 3.0 & 1.3\\
171003 & 842.3 & 12.4 & 13.0 & 9.0 & 1.0 \\
171004 & 949.0 & 12.6 & 16.6 & 9.8 & 1.0 \\
171019 & 485.7 & 0.2 & 12.4 & 9.7 & 0.9 \\
171020 & 1148.4 & 3.6 & 4.5 & 2.3 & 1.1 \\
171116 & 1331.9 & 1.0 & 4.0 & 3.3 & 0.7 \\
171213 & 965.3 & 0.0 & 4.0 & 3.8 & 0.3 \\
171216 & 205.7 & 2.1 & 1.3 & 0.0 & 0.0 \\
180110 & 1338.9 & 6.0 & 4.5 & 3.4 & 1.3 \\
180119 & 965.3 & 0.0 & 3.9 & 5.7 & 0.3 \\
180128.0 & 801.3 & 8.5 & 7.0 & 7.8 & 1.7 \\
180128.2 & 343.4 & 0.8 & 2.8 & 0.0 & 0.0 \\
180130 & 1338.9 & 6.0 & 3.4 & 3.7 & 1.3 \\
180131 & 912.9 & 1.5 & 4.4 & 3.8 & 1.3 \\
180212 & 783.4 & 6.8 & 7.9 & 7.4 & 1.2 \\
180315 & 60.0 & 0.0 & 6.6 & 2.7 & 1.3 \\
180324 & 49.0 & 1.4 & 3.6 & 7.0 & 1.0 \\
180417 & 0.0 & 0.0 & 6.7 & 0.0 & 0.0 \\
180430 & 10.9 & 0.0 & 1.0 & 8.3 & 1.4 \\
180515 & 3.0 & 0.0 & 4.1 & 2.7 & 0.0 \\
180525 & 737.8 & 9.8 & 4.8 & 5.1 & 2.0 \\
180924 & 912.9 & 1.5 & 10.8 & 4.0 & 1.3 \\
\hline
Total & 18,162 & 90.6 & 238.7 & 118.4 & 26.1 \\
	\hline
	\end{tabular}
	\caption{Total time of observations (hr) per telescope and target FRB.
	As some FRBs occurred in the same ASKAP field, some time spent observing with ASKAP covered multiple FRBs at once. See the caption of Table~\ref{tab:telescope_properties} for definitions of acronyms.
	}
	\label{tab:observations_sum}
\end{table}

In a total of 383.2\,hr of follow-up time (i.e.\ on instruments other than ASKAP), two repeat bursts, and one new FRB, were detected. The repeat bursts, both from FRB~171019, are reported elsewhere \citep{Kumar2019}, and the analysis of the new FRB is ongoing. Our key result, however, is that in this large follow-up campaign, and during ASKAP observations of the same field, only one FRB was detected to repeat. We thus proceed to derive limits on the repetition rates of each object, should they indeed repeat at all.

\section{Model of repeating FRBs}
\label{sec:model}

We base our model of a repeating FRB on the behaviour of FRB~121102. \repeater\ resides in a dwarf galaxy host at a redshift of $z \approx 0.19$  \citep{Chatterjee2017,Tendulkar2017}. The bursts have DMs consistent with arising from a constant DM of 559.6\,pc\,cm$^{-3}$, with an observational scatter of 4.2\pccc\ per burst \citep{Hardy2017}. The host galaxy has an apparent $r$ band magnitude of $m_r = 25.1$ AB mag, and a stellar mass of $4$-$7 \times 10^7$ M$_\odot$ \citep{Tendulkar2017}. The H$_\alpha$ flux of the galaxy indicates a substantial contribution to the burst's DM (due to the host) of  up to 324\pccc\ \citep{Tendulkar2017}.

Since only one of the 27 ASKAP FRBs used in this analysis has been localised to its host galaxy, the model is written in terms of observable properties: fluence $F$, and rate in the observer frame $R$, at the mean frequency $\overline{\nu}=1.296 \approx 1.3$\,GHz at which most were discovered. We discuss the implications for the intrinsic properties of these sources in Section~\ref{sec:absolute_rates}.

\subsection{Fluence distribution}
\label{sec:fluence_distribution}

Given the range of telescope sensitivities used to observe CRAFT FRBs, and the variation in source distance expected from their DMs, the burst fluence distribution strongly affects the relative rates at which each telescope should detect repeat bursts. \citet{Law2017} presented a study of 17 repeated bursts from  \repeater, estimating the cumulative slope of this distribution in log-log space to be $\gamma = -0.7$, with a burst rate above $10^{38}$\,erg of approximately once per hour. We therefore describe the cumulative burst rate distribution $R$ as
\begin{eqnarray}
R(F_{\rm 1.3\,GHz}) & = & R_0 \left( \frac{F_{\rm 1.3\,GHz}}{1\,{\rm Jy\,ms}} \right)^\gamma, \label{eq:rate_scaling}
\end{eqnarray}
where $F_{\rm 1.3\,GHz}$ is the fluence at 1.3\,GHz, $R_0$ is the rate of bursts with fluence above 1\,Jy\,ms, and $\gamma$ is the cumulative power-law index.

\citet{Jamesetal2019a} have recently shown that $s$, being the ratio between the signal-to-noise ratio (S/N) of each burst, and the threshold S/N used in the detection algorithm, S/N$_{\rm th}$, will follow the same power-law as the true underlying fluence distribution. This allows all data on \repeater\ where these values have been published to be used to estimate $\gamma$. Applying this method to the nine VLA bursts from \citet{Law2017}, and 21 bursts from \citet{Gajjaretal2018}, \citet{James2019b} finds $\gamma=-0.91 \pm 0.17$.

\citet{Gourdjietal2019}, using 29 bursts from a total sample of 41 bursts detected by Arecibo, estimate $\gamma=-1.8 \pm 0.3$. Using the $s$ statistic, which allows the inclusion of all bursts while reducing potential sources of bias, produces $\gamma=-2.3_{-0.3}^{+0.4}$. This is clearly in conflict with previous results. It is also internally inconsistent with a power-law: the distribution of S/N within the sample is extremely peaked towards near-threshold events.

A population of FRBs repeating similarly to \repeater --- i.e.\ as per equation~(\ref{eq:rate_scaling}) --- will produce an intrinsic luminosity distribution for the FRB population with the same power-law index $\gamma$. \citet{Macquart2018b} note that the FRB population must exhibit a burst strength index flatter than $\gamma=-1.5$ in order to obtain a cosmological distribution of bursts dominated by the intrinsically brightest events, as now found for the ASKAP/CRAFT sample by \citet{Shannonetal2018}. \citet{Lu2019} find $\gamma=-0.6 \pm 0.3$ for the intrinsic luminosity distribution of the ASKAP/CRAFT FRBs, and \citet{2016MNRAS.461L.122L} find $-1.2 \le \gamma \le -0.5$ for Parkes data.

In this work, we therefore consider the range $\gamma = -1 \pm 0.5$ to cover a broad range of possible burst strength indices, while noting the range $\gamma=-0.9 \pm 0.2$ is most likely.

\subsection{Spectral properties}
\label{sec:spectral_properties}

Bursts observed from \repeater\ are contained in a relatively narrow frequency range  \citep{Law2017}, and are typically composed of several temporal sub-bursts \citep{Hessels2019}. As all searches used in this work use the entire bandpass with equal weighting to evaluate burst S/N, spectral structure on frequency scales much smaller than the bandpass will not affect telescope sensitivity. However, the different observation frequencies and bandwidths require a model for how the burst rate scales between instruments.

To develop such a model, we use a power-law with spectral index $\alpha$, such that the fluence at frequency $\nu$ is
\begin{eqnarray}
F(\nu) & = & F_{\rm 1.3\,GHz} \left( \frac{\nu}{1.3\,{\rm GHz}} \right)^\alpha. \label{eq:frequency_scaling}
\end{eqnarray}
We do not consider any low-frequency cut-off in the burst spectrum, as suggested by \citet{Sokolowskietal2018}, since the CHIME collaboration have observed bursts down to 400\,MHz \citep{CHIME2019a}, below the frequency ranges of the observations reported here.

\citet{Macquart2019a}, analysing a sample of 23 ASKAP bursts almost identical to that used here, find $\alpha=-1.5_{-0.3}^{+0.2}$ from the distribution of spectral power within the 336\,MHz wide ASKAP band. For FRB~171019, \citet{Kumar2019} find evidence for a much steeper spectral index than the nominal value of $\alpha=-1.5$, with most-likely values near $\alpha=-8$ or steeper. This cannot be typical of all ASKAP FRBs, nor typical of the population observed by CHIME down to 400\,MHz, since far more bursts would then have been detected. It is also inconsistent with the non-detections of ASKAP FRBs by the Murchison Widefield Array \citep{Sokolowskietal2018}. However, it is possible for a single, unusual object to have properties very different from that of the entire population. Therefore, we assume $\alpha=-1.5$ for the majority of FRBs in Section~\ref{sec:limits}, and consider $0 \le \alpha \le -8$ for FRB~171019 in Section~\ref{sec:FRB171019}.

Equation~(\ref{eq:frequency_scaling}) can be interpreted as either modelling the  spectral index of individual broadband bursts, or as modelling the frequency-dependent rate of bursts individually contained within a narrow bandwidth through equation~(\ref{eq:rate_scaling}). The fluence thresholds $F_{\rm th}$ quoted in  Table~\ref{tab:telescope_properties} are calculated assuming full band occupancy. Should observational bandwidth increase beyond the characteristic bandwidth of repeat bursts however, both the fluence threshold $F_{\rm th}$, and number of bursts occurring within the bandwidth, will increase linearly. The former effect will act to decrease the observation rate, while the latter will increase it. For the fluence dependence given by equation~(\ref{eq:rate_scaling}), the total rate
\begin{eqnarray}
R & \propto & \left( \frac{\Delta \nu}{336\,{\rm MHz}} \right)^{\gamma+1}, \label{eq:band_occupancy}
\end{eqnarray}
where the standard ASKAP bandwidth of $336$\,MHz is used as a normalisation constant. Given that $\gamma$ is likely to be in the range $-0.5$ to $-1.5$ (see Section~\ref{sec:spectral_properties}), the total rate will scale with bandwidth to the power of $\pm0.5$.

For computational simplicity, we only consider two cases in this work. The `standard' case, using the nominal thresholds of Table~\ref{tab:telescope_properties} regardless of bandwidth, applies to broadband bursts, or to narrow-band bursts when $\gamma=-1$ through equation~(\ref{eq:band_occupancy}).
We also consider a case where
\begin{eqnarray}
F_{\rm th} & \propto & \left( \frac{\Delta \nu}{336\,{\rm MHz}} \right)^{0.5},
\end{eqnarray}
which is equivalent to $\gamma=-2$. As noted in Section~\ref{sec:fluence_distribution}, this is disfavoured by current measurements, i.e.\ this scenario is perhaps overly pessimistic. This scenario is termed `low band occupancy'. It represents a burst occupying a small range of frequencies in the observation band, reducing the event rate with bandwidth when $\gamma=-2$.  The `standard' scenario also represents low band occupancy when $\gamma=1$.

\subsection{Arrival time distribution}

The arrival-time distribution of bursts from \repeater\ appears to be clustered, with bursts typically discovered in groups \citep[e.g.][]{Gajjaretal2018}. \citet{Oppermannetal2018} find that a Weibull distribution describes the observed clustering of bursts, and measure a repeat rate of $6^{+3}_{-2}$ events per day for fluences $>20$ mJy, and a clustering parameter $k$ of $0.34^{+0.06}_{-0.05}$.

The Weibull distribution is commonly used in failure analysis. In this context, a value of $k<1$ models cases where failure is likely to occur immediately (e.g.\ due to defects), with a failure rate decreasing over time, while $k>1$ models cases where the failure probability increases with time, e.g.\ due to ageing. The case of $k=1$ is a failure rate independent of time, i.e.\ a Poisson process.
In the context of emission from a repeating FRB, $k<1$ indicates a clustered distribution, with a high probability of short wait times between bursts, but also a high probability of long periods of inactivity, while $k=1$ is the Poisson case with an exponential wait time distribution between bursts. Note that this is in contrast to a model with active and inactive periods, with the former having a higher emission rate than the latter, but with burst times being governed by a Poisson process within each period.

We model the potentially clustered nature of repeating FRBs using the same approach as \citet{Oppermannetal2018}, i.e., the probability distribution $P$ of wait times $\delta t$ between successive bursts
\begin{eqnarray}
P(\delta t | k,R) & = & \frac{k}{\delta t} \left[ \delta t \, R \, \Gamma(1+k^{-1}) \right]^k e^{-\left[ \delta t\, R \,\Gamma(1+k^{-1}) \right]^k}, \label{eq:wait_times}
\end{eqnarray}
where $\Gamma$ is the gamma function. This parametrization holds $R$ constant while changing $k$. For completeness, we investigate the range $k=$0.1--1.

\subsection{Burst width and scattering}

The sensitivity of an FRB search reduces with burst width, as a finite fluence is spread over more noise. The burst width is attributable to an intrinsic width, which we assume is frequency independent; scatter broadening, which will increase at lower frequencies as $\sim \nu^{-4.4}$; and dispersion smearing, due to the finite width of each frequency channel. The intrinsic durations of bursts (or groups of sub-bursts) from \repeater\ are 1--5\,ms \citep{Hessels2019}, with sub-burst structure down to $0.1$\,ms. The bursts are not significantly affected by scattering.

The durations of bursts measured by ASKAP vary from approximately the 1.27\,ms time resolution of the search to 5\,ms, with a scattering tail detectable in the sample of \citet{Shannonetal2018} only for the brightest burst, FRB~180110. Here, we use the detected widths of ASKAP bursts to estimate telescope sensitivity at 1.3\,GHz, and consider different sets of assumptions in scaling to other frequencies.

The three considered assumptions are that the burst widths observed by ASKAP contain no scattering contribution; that they are entirely scatter-dominated; and (the most pessimistic case) that they are scatter dominated at frequencies below $1.3$\,GHz, but limited by their intrinsic width at higher frequencies. In these scenarios, the observed frequency-dependent burst width $w_{\rm obs}(\nu)$ is scaled from the ASKAP width $w_{\rm A}$ as
\begin{eqnarray}
w_{\rm obs}(\nu) = \begin{cases} w_{\rm A} & {\rm no~scattering} \\
        w_{\rm A} \left(\frac{\nu}{1.3\,{\rm GHz}}\right)^{-4.4} & {\rm scattering}  \\
       w_{\rm A} \, {\rm MAX}\left[ 1, \left(\frac{\nu}{1.3\,{\rm GHz}}\right)^{-4.4}\right] & {\rm pessimistic}
        \end{cases}. \label{eq:burst_width}
\end{eqnarray}
The width due to dispersion smearing within each frequency channel, $w_{\rm DM}$, is given by
\begin{eqnarray}
w_{\rm DM} & = & 8.3 \,{\rm \mu s} \, \frac{\rm DM}{1\,{\rm pc}\,{\rm cm}^{-3}} \frac{\delta \nu}{1\, \rm MHz}  \left( \frac{\nu}{1\, \rm GHz} \right)^{-3},
\end{eqnarray}
where $\delta \nu$ is the channel width from Table~\ref{tab:telescope_properties}.

\subsection{Sensitivity dependence}
\label{sec:sensitivity_dependence}

For a given fluence, the sensitivity to a transient source scales with its effective duration $w_{\rm eff}$ as
\begin{eqnarray}
F_{\rm th} & = & F_{\rm th}(1\,{\rm ms}) \sqrt{\frac{w_{\rm eff}}{1\,{\rm ms}}}. \label{eq:width_sensitivity}
\end{eqnarray}
We model the total effective width of a burst, $W_{\rm eff}$, following \citet{Cordes_McLaughlin_2003}, using the geometric sum of its individual widths
\begin{eqnarray}
w_{\rm eff} & = & \sqrt{w_{\rm DM}^2 + w_{\rm samp}^2 + w_{\rm obs}^2}, \label{eq:effective_width}
\end{eqnarray}
where $w_{\rm DM}$ is the smearing of dispersion measure in each frequency channel (evaluated at band centre), $w_{\rm samp}$ is the time resolution used for the (incoherent) dedispersion search, and $w_{\rm obs}$ is given by equation~(\ref{eq:burst_width}).

\subsection{Implementation in a simulation}

Limits on the repetition properties are generated as follows. Each simulation run is characterised via the parameter set $k$, $R$, $\gamma$, a set of assumptions on band occupancy, burst width, and spectral index $\alpha$, and the FRB in question.

\begin{table}
    \centering
    \begin{tabular}{c | cc |c c| c}
  $F_{\rm eff}$     & \multicolumn{2}{|c|}{ASKAP}  & Parkes & \multicolumn{2}{c|}{GBT} \\
Assumption       &   FE & ICS & MB & 820\,MHz & L-band \\
\hline
Standard & 64 & 13 & 0.82 & 0.28 & 0.11\\
Flat spectrum & 64 & 13.5  & 0.75 & 0.56 & 0.87\\
No scattering & 64 & 12.5 & 0.95  & 0.10 & 0.13 \\
Pessimistic Scat. & 64 & 13 & 0.95  & 0.28 & 0.13 \\
Low band occupancy & 64 & 13  & 0.83  & 0.28 & 0.17 \\
    \end{tabular}
    \caption{Effective telescope thresholds (Jy\,ms), to FRB~180128.0 (DM $=441.7$\,pc\,cm$^{-3}$, $w_{\rm A}=2.9$\,ms), i.e.\ telescope thresholds are scaled relative to ASKAP Fly's Eye observation parameters from Table~\ref{tab:telescope_properties}, under different sets of assumptions. These are the `standard' scenario ($\alpha=-1.5$, $w \propto \nu^{4.4}$, full band occupancy); setting a flat spectrum ($\alpha=0$); no scattering ($w=w_{\rm A}$), pessimistic scattering from equation~(\ref{eq:burst_width}); and low band occupancy from equation~(\ref{eq:band_occupancy}). The values for ASKAP ICS mode are typical examples. See Table~\ref{tab:telescope_properties} for definitions of acronyms.}
    \label{tab:effective_thresholds}
\end{table}

The list of observations for that FRB for each telescope is loaded and sorted in chronological order.  The nominal thresholds are then scaled to effective thresholds at 1.3\,GHz using the observed width $w$ and DM of each burst, and each observation's instrumental and detection parameters from Table~\ref{tab:telescope_properties}. In the case of ASKAP ICS observations, the frequency, bandwidth, and number of telescopes $N$ varied for each observation, and the effective threshold is calculated accordingly, using $F_{\rm th}\sim N^{-0.5}$ \citep{2019Sci...365..565B}.
The sensitivity of each ASKAP observation is also scaled according to the sensitivity of the discovery beam, and position in that beam, according to \citet{2019PASA...36....9J}. Due to an error in metadata, some observations with the Parkes multibeam were offset from the position of the FRB being followed-up. In these cases, the telescope threshold is increased to account for the reduced sensitivity away from beam centre. For most Parkes and all GBT observations however, the location of the FRB was sufficiently well-localised \citep[see ][]{Shannonetal2018} that no beam correction is needed.

We use the central frequency only to characterise telescope sensitivity --- for burst spectral indices in the range $-2 \le \alpha \le 0$, this leads to errors of less than 10\% in assumed sensitivity for the GBT L-band receiver, and less than 1\% for the other instruments. For the investigated range of $\gamma$, corresponding rate errors will be comparable. Since the GBT L-band receiver contributed only 7\% of the total follow-up time, total expected rate errors will be at the 1\% level.

An example of the effective thresholds $F_{\rm eff}$ calculated with this procedure are given in Table~\ref{tab:effective_thresholds}, for different sets of assumptions on burst width, band occupancy, and spectral index. By design, the sensitivity of ASKAP Fly's Eye observations is unaffected by this choice of assumption, since $F_{\rm eff}$ is normalised by these observing parameters. The variation in ASKAP ICS observations is due to some observations having lower frequency and/or a longer integration time. The Parkes multibeam, with similar bandwidth and observation frequency to ASKAP, also has an approximately constant $F_{\rm eff}$. The telescope most affected is the GBT. Observations with the 800\,MHz receiver are at a lower frequency, and L-band observations have a much broader bandwidth. This results in sensitivity varying by factors of 5--10 between different assumptions.

\begin{figure}
    \centering
    \includegraphics[width=\columnwidth]{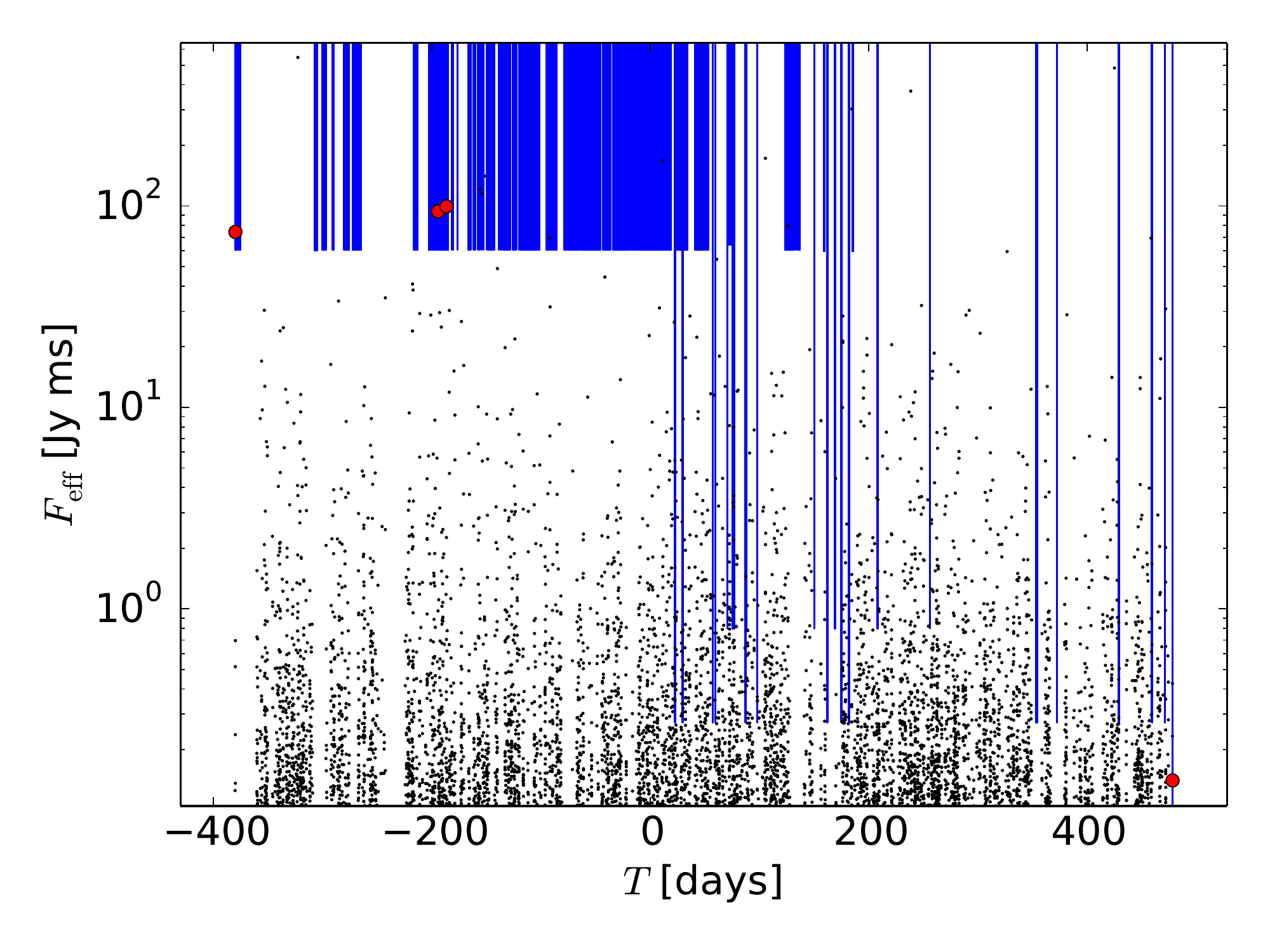}
    \caption{Illustration of the simulation of a sequence of bursts (black dots), compared to observations (blue lines), and detections (red circles). The observation lines have minima at the effective threshold, and span the time range observed. This case is for FRB~180119, for $R_0=1$\,day$^{-1}$, $k=0.34$, $\gamma=-0.9$, and the standard set of assumptions. The original discovery, at time $T=0$ days, is not shown. Note that many ASKAP observations are sufficiently close in time that the lines overlap.}
    \label{fig:example_simulation}
\end{figure}

\subsection{Simulation method}
\label{sec:simulation_method}

Beginning with the time of the FRB discovery, sequences of burst wait times are drawn according to equation~(\ref{eq:wait_times}), using the Weibull parameter $k$, and the rate $R$ scaled from a nominal value above 1\,Jy\,ms to the lowest value of effective telescope threshold. Time is defined relative to the initial discovery, and sequences must begin at that time. The Weibull distribution is statistically identical when generating bursts both forwards and backwards in time, and this is done to cover all observations that would have been sensitive to that FRB, i.e.\ both before and after the initial discovery. The fluence of each burst generated during an observation period is sampled according to the differential power-law index of $\gamma-1$. If that burst passes the telescope threshold, it is counted as a detection (the initial discovery is ignored). An example of such a sequence is shown in Fig.~\ref{fig:example_simulation}.

Discounting the initial discovery is a critical statistical step in our analysis. Since we do not estimate the population of repeating FRBs from which we observe no bursts at all, using the initial detections would create a bias towards high burst rates. Rather, these are used to identify the presence of a potentially repeating FRB, and we model the probability of a repeat burst given the time of the initial observation.

\begin{figure}
    \centering
    \includegraphics[width=\columnwidth]{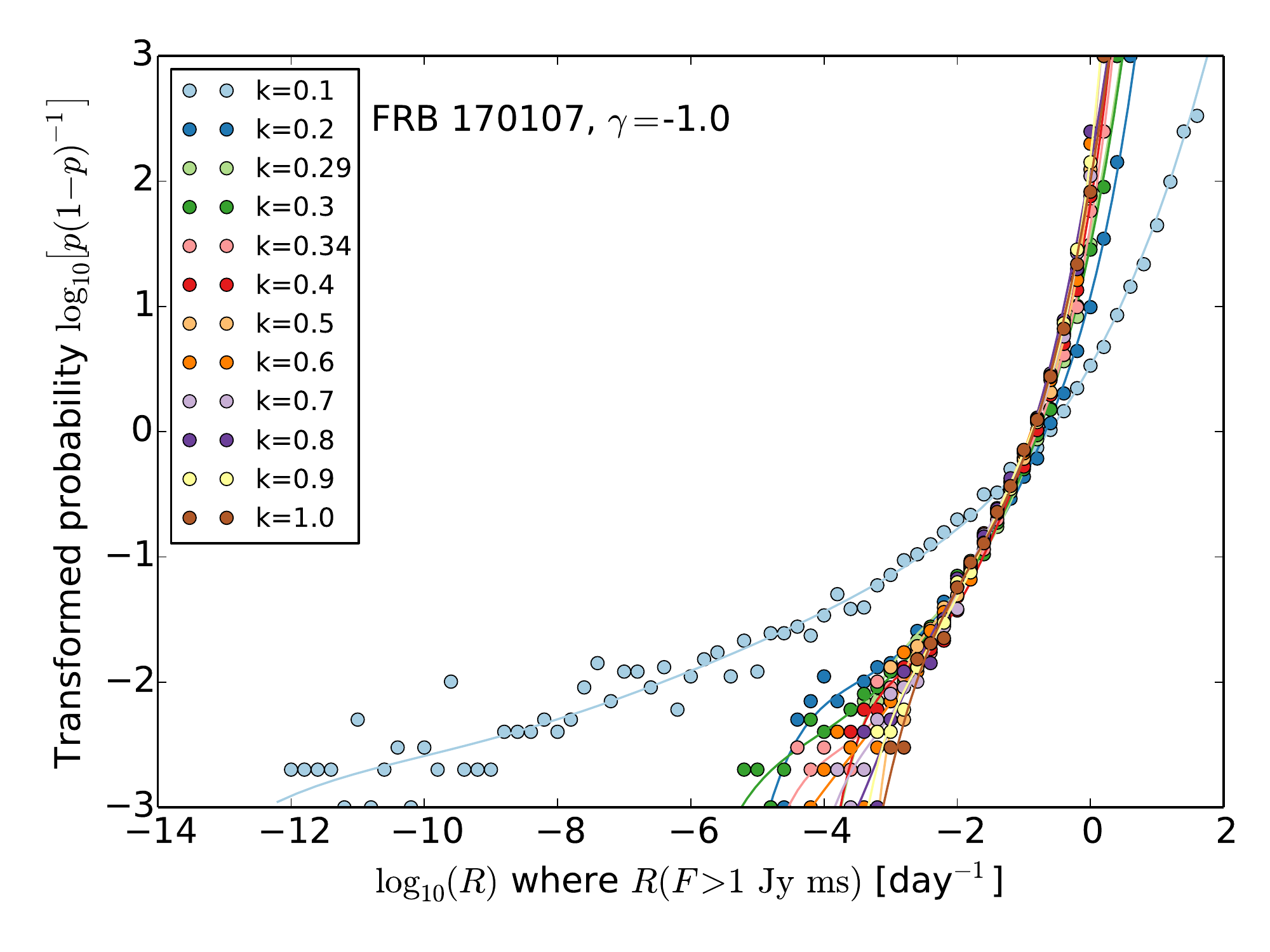}
    \includegraphics[width=\columnwidth]{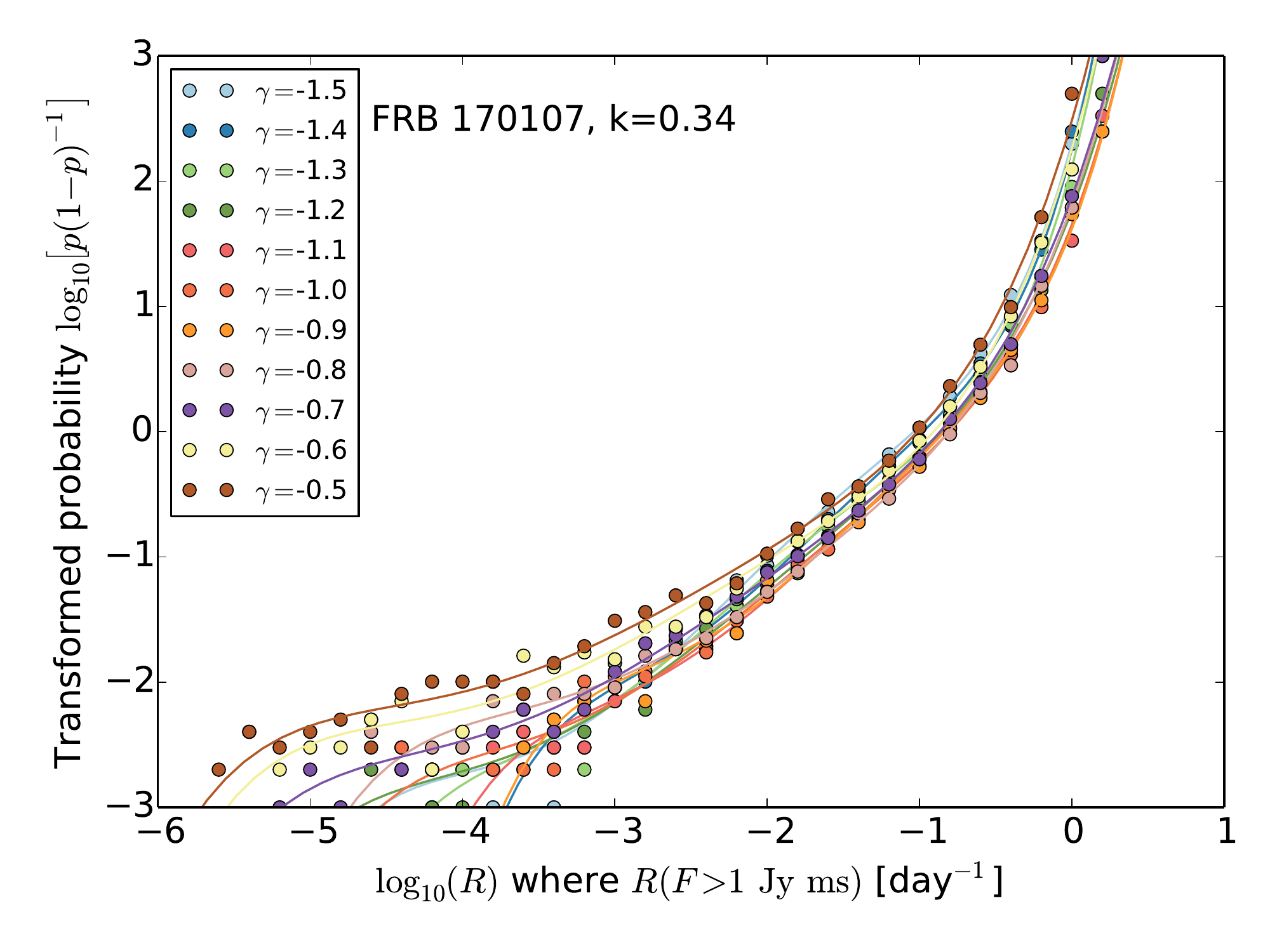}
    \caption{Probability $p(R|\gamma,k)$ of detecting one or more repeat bursts, as a function of burst rate $R$ above 1\,Jy\,ms, for FRB~170107. Top: varying $k$ for fixed $\gamma=$-0.9; bottom: varying $\gamma$ for fixed $k=0.34$. Points are simulated probabilities, lines are fifth-order polynomial fits.}
    \label{fig:p_fit}
\end{figure}

In order to obtain a good statistical estimate of the probability of detecting a repeat burst, 1000 such sequences are generated for each simulation run. The total number of sequences in which one or more repeat bursts are detected is recorded. The rate $R$ is then increased until all 1000 such sequences produce more repeat bursts than observed for that FRB, and reduced until no repetitions are observed. These data are then used to fit probabilities $p(R|k,\gamma)$ of any given repetition outcome (e.g.\ no detected repeats) as a function of $R$ for each $k$, $\gamma$, and set of assumptions. An example of these fits --- performed with SciPy \citep{SciPy2019}, with a $5^{\rm th}$-order polynomial in $\log R$--$\log p/(1-p)$ space to obtain smooth results in both the limits $p\to0$ and $p\to1$ --- is given in Fig.~\ref{fig:p_fit}. The fits are reliable in the range $0.002<p<1$, allowing limits of up to 99.7\% ($3 \sigma$) confidence to be set.

\section{Results}

\subsection{Limits on repetition for FRBs with no repeats}
\label{sec:limits}

\begin{figure}
    \centering
    \includegraphics[width=\columnwidth,clip=true, trim={0.8cm 0cm 1.2cm 1cm}]{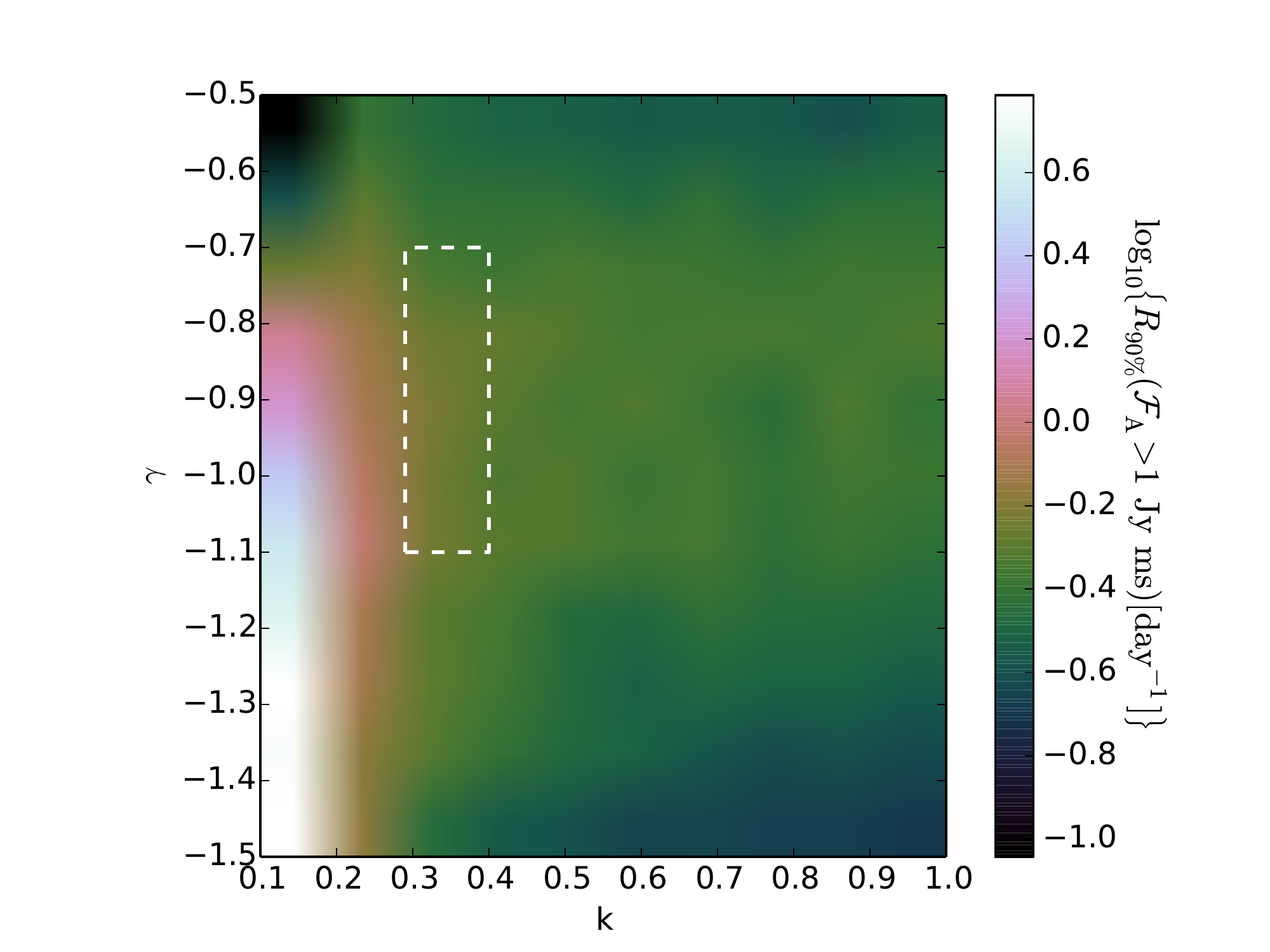}
    \caption{Upper limits at 90\% confidence, $R_{90}$, on the repetition rate $R$ of bursts above 1\,Jy\,ms from FRB~170107 as a function of $k$ and $\gamma$. The white dashed box indicates the ranges on these parameters for FRB~121102 --- see Section~\ref{sec:model}.}
    \label{fig:contour_plot}
\end{figure}

For those FRBs with no observed repeat bursts (all except FRB~171019), confidence limits on repetition rates $R$ can be set when the probability of detecting one or more repeat bursts is equal to the desired level of confidence. Tests showed that quoting limits at the threshold of 1\,Jy\,ms showed least variation with $\gamma$. Fig.~\ref{fig:contour_plot} displays the 90\% confidence upper limits, $R_{90}$, on the burst rate $R$ above 1\,Jy\,ms for FRB~170107. For the standard scenario ($\alpha=-1.5$, scattering as $\nu^{-4.4}$, full band occupancy, $k=0.34$, $\gamma=-0.9$), $R_{90}$ was found to be 0.56\,day$^{-1}$. Holding $k$ and $\gamma$ constant and varying assumptions regarding scattering, band occupancy, and allowing spectral index to be $0$, varied $R_{90}$ between 0.43\,day$^{-1}$ and 0.60\,day$^{-1}$, i.e.\ ${}_{-25}^{+3}$\%. This is comparable to the variation when allowing only $k$ and $\gamma$ to fluctuate within their nominal ranges of $0.29$ to $0.4$ and $-0.7$ to $-1.1$ respectively ($R_{90}$ varying by ${}_{-17}^{+13}$\%). Allowing very clustered distributions ($k=0.1$) results in weaker limits for $\gamma < -0.9$, with $R_{90}=6$\,day$^{-1}$ for $\gamma=-1.5$. This is because more-negative values of $\gamma$ place emphasis on the more sporadic follow-up observations with Parkes and GBT, which could easily miss an outburst. The strongest limits of $R_{90}=0.09$\,day$^{-1}$ are placed for $\gamma=-0.5$, $k=0.1$, since no secondary bursts were detected by ASKAP close to the initial detection.

\begin{table}
    \centering
    \begin{tabular}{c| c | c c | c c}
       $k$ & 0.34 & \multicolumn{2}{|c}{$0.29\le k \le 0.4$} & \multicolumn{2}{|c}{$0.1 \le k\le 1$} \\
       $\gamma$ & -0.9 & \multicolumn{2}{|c}{$-0.7\le \gamma \le -1.1$} & \multicolumn{2}{|c}{$-0.5\le \gamma \le -1.5$} \\
       FRB  & $R_{90}$ & $R_{90}^{\rm min}$ & $R_{90}^{\rm max}$& $R_{90}^{\rm min}$ & $R_{90}^{\rm max}$ \\
       \hline
  170107 & 0.56 & 0.5 & 0.63 & 0.091 & 6.1\\ 
  170416 & 3.1 & 2.4 & 5.3 & 0.095 & 67\\ 
  170428 & 1.3 & 0.95 & 1.8 & 0.38 & 16\\ 
  170707 & 8.4 & 5.3 & 16 & 1.5 & 140\\ 
  170712 & 6.3 & 5.4 & 12 & 2.0 & 130\\ 
  170906 & 1.2 & 0.97 & 1.7 & 0.03 & 10\\ 
  171003 & 0.96 & 0.82 & 1.2 & 0.046 & 9.4\\ 
  171004 & 0.66 & 0.6 & 0.83 & 0.026 & 4.0\\
  171020 & 2.4 & 2.1 & 2.8 & 0.15 & 37\\ 
  171116 & 1.5 & 1.0 & 2.4 & 0.023 & 14\\ 
  171213 & 1.8 & 1.4 & 2.1 & 0.056 & 20\\ 
  171216 & 13 & 8.3 & 31 & 0.14 & 200\\ 
  180110 & 1.9 & 1.4 & 2.8 & 0.038 & 23\\ 
  180119 & 1.2 & 0.92 & 1.7 & 0.026 & 10\\ 
  180128.0 & 1.2 & 0.97 & 1.4 & 0.046 & 11\\ 
  180128.2 & 6.4 & 4.5 & 15 & 0.029 & 78\\ 
  180130 & 1.7 & 1.2 & 2.4 & 0.029 & 18\\ 
  180131 & 1.9 & 1.6 & 2.9 & 0.042 & 24\\ 
  180212 & 1.2 & 1.0 & 1.4 & 0.089 & 10\\ 
  180315 & 2.9 & 2.5 & 3.8 & 0.55 & 47\\ 
  180324 & 3.1 & 2.4 & 3.7 & 0.85 & 57\\ 
  180417 & 18 & 14 & 23 & 4.7 & 2400\\ 
  180430 & 1.7 & 1.1 & 2.6 & 0.3 & 47\\ 
  180515 & 6.9 & 5.8 & 9.7 & 1.2 & 190\\ 
  180525 & 2.6 & 2.3 & 3.9 & 0.62 & 49\\ 
  180924 & 2.0 & 1.6 & 3.1 & 0.46 & 110\\ 
    \end{tabular}
    \caption{Upper limits at 90\% confidence, $R_{90}$, on the repetition rate $R$ above 1\,Jy\,ms for each FRB for different ranges of Weibull index $k$ and burst fluence index $\gamma$, assuming the standard parameter set with $\nu^{-4.4}$ scattering, spectral index $\alpha=-1.5$, and full band occupancy. These have estimated systematic uncertainties of $\pm25\%$ due to different assumptions regarding band occupancy, burst width, and spectral index $\alpha$.}
    \label{tab:upper_limits}
\end{table}

Given the dominance of uncertainties in $k$ and $\gamma$, for other FRBs in the sample, we only simulate the `standard' scenario. Table \ref{tab:upper_limits} reports 90\% confidence upper limits, $R_{90}$, on the repetition rate above 1\,Jy\,ms, and the variation in $R_{90}$ when varying ($k,\gamma$) over the ranges ($0.29$ to $0.34$,$-1.1$ to $-0.7$), and ($0.1$ to $1$,$-1.5$ to $-0.5$), respectively. The differences in limits between FRBs reflect the integration times from Table~\ref{tab:observations_sum} and telescope sensitivities from Table~\ref{tab:telescope_properties}.

\subsection{Limits on repetition for FRB~171019}
\label{sec:FRB171019}

In the case of FRB~171019, limits on burst properties can be derived considering that precisely two repeat bursts were observed; that these bursts were observed by the GBT 800\,MHz receiver; and that the bursts were observed on 2018-07-20 at 08:33:37\,UT (observation of 1200\,s duration) and 2019-06-09 at 07:40:46\,UT (observation of 8397\,s duration). The first piece of information allows both upper and lower rate limits to be set for any given $k$, $\gamma$, and $\alpha$. It also disfavours small values of $k$ distributions, since such clustered distributions tend to produce either no or many repeats.

The second clearly constrains the valid range of $\gamma$ and $\alpha$, and will favour steep spectral indices for both, since the bursts were observed in the lowest-frequency observation only, and the far more numerous ASKAP Fly's Eye observations made at only slightly higher frequencies were at a lower sensitivity.

The third also disfavours clustered burst arrival times, since these would likely be observed in the same observation period.

We simulate burst sequences for FRB~171019 as per Section~\ref{sec:simulation_method} as a grid in $k$--$\gamma$--$\alpha$, recording the fraction of bursts satisfying the above criteria. Since it is computationally intensive to recreate the exact observation times of repeat bursts, we count all instances where two bursts are detected by the GBT at 800\,MHz in two different observations as satisfying our constraints. We again fit the simulated probability as a function of $R$, $P(R|k,\gamma,\alpha)$, and use its maximum, $P_{\rm max}(k,\gamma,\alpha)$, to set confidence limits.

\begin{figure}
    \centering
    \includegraphics[width=\columnwidth,clip=true,trim={0cm 1cm 0cm 3cm}]{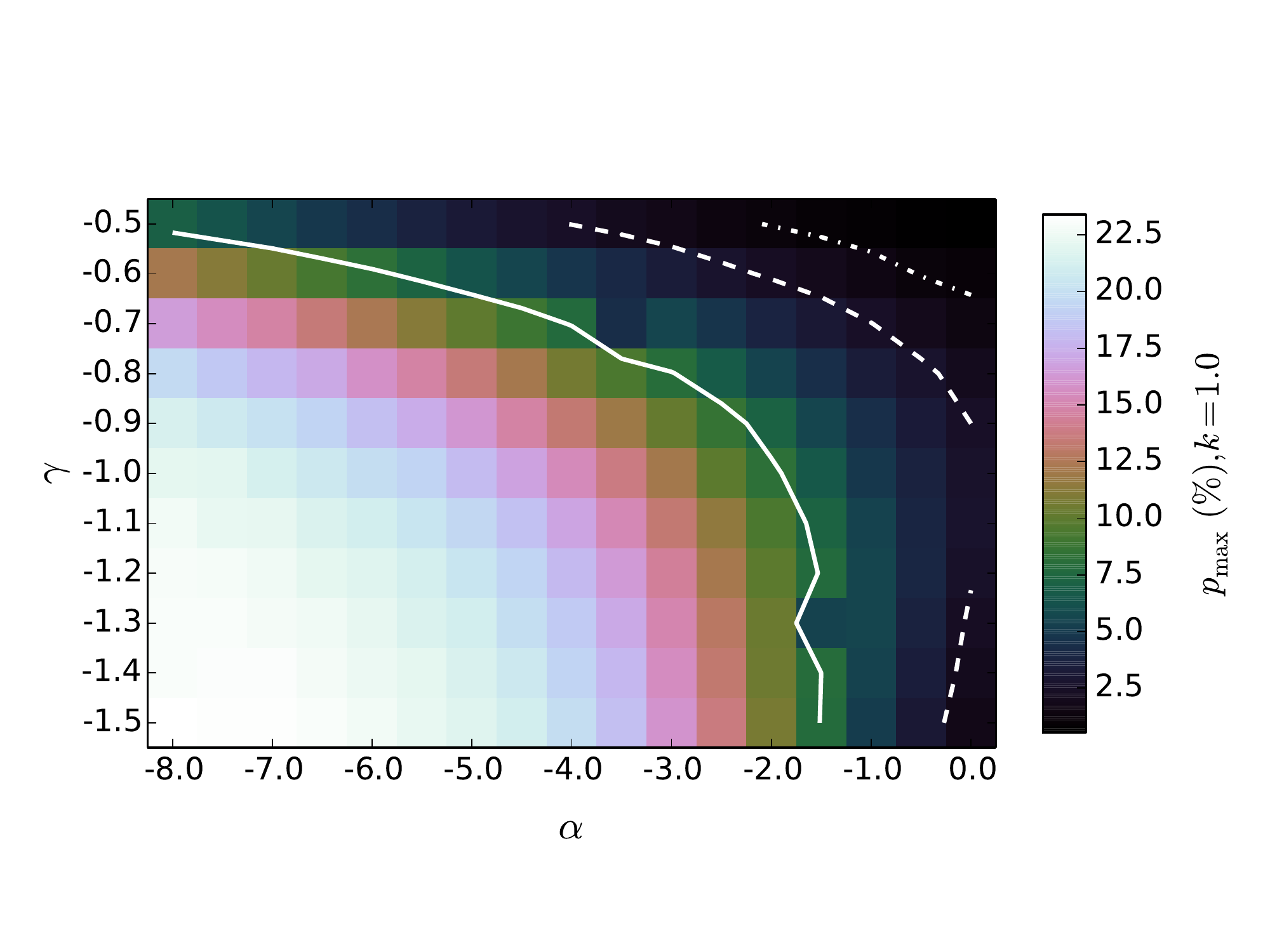}
    \includegraphics[width=\columnwidth,clip=true,trim={0cm 1cm 0cm 3cm}]{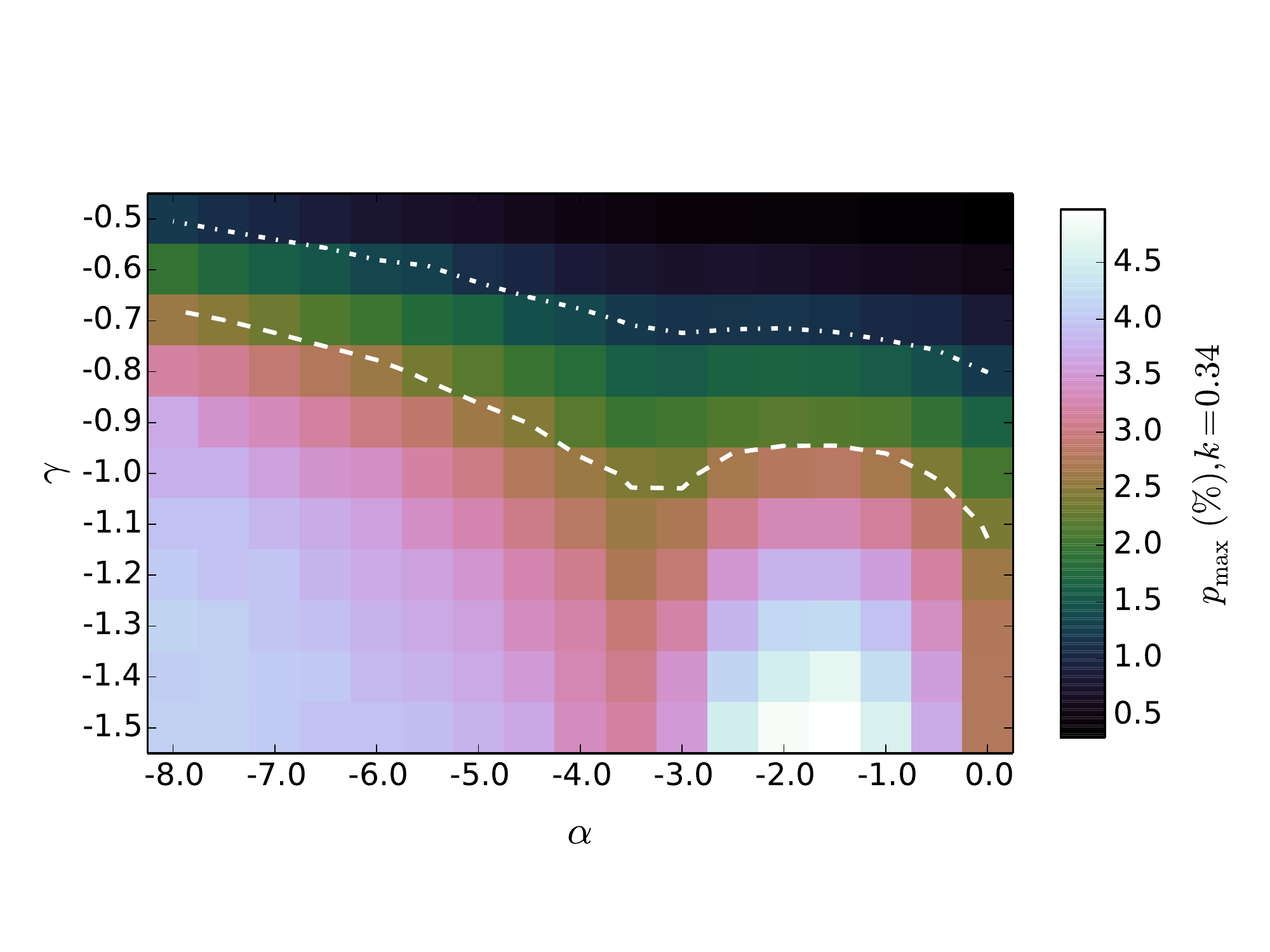}
    \includegraphics[width=\columnwidth,clip=true,trim={0cm 1cm 0cm 3cm}]{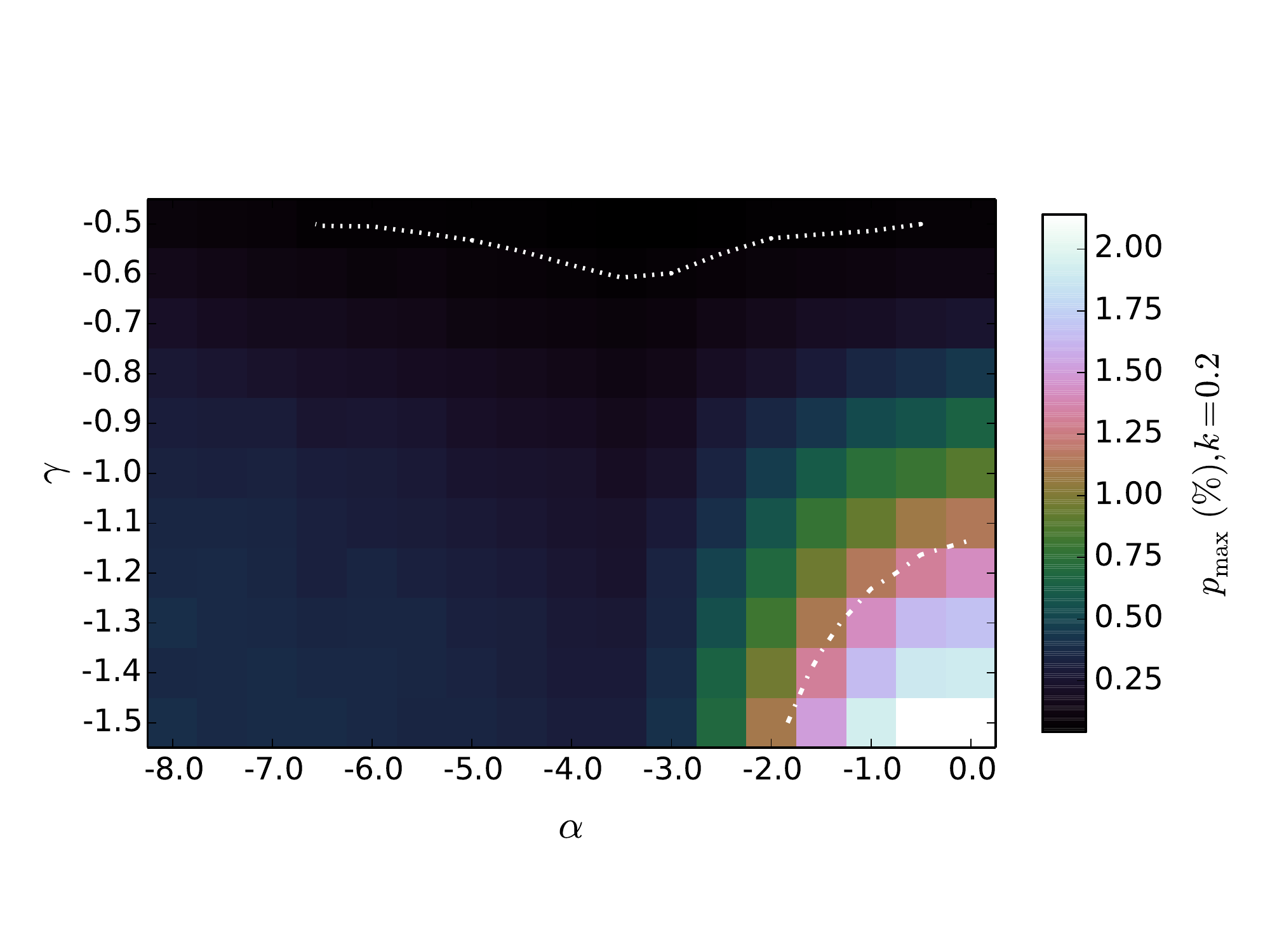}
    \caption{Simulated probability of CRAFT follow-up observations of FRB~171019 after marginalisation over $R$, $p_{\rm max}(\alpha,\gamma)$, for three values of Weibull burst index: $k=1$ (top), $k=0.34$ (middle), and $k=0.2$ (bottom). The solid, dashed, dot-dash, and dotted contours respectively indicate 68\%, 90\%, 95\%, and 99.7\% exclusion regions.}
    \label{fig:caseC}
\end{figure}

Fig.~\ref{fig:caseC} shows $P_{\rm max}(k,\gamma,\alpha)$ for three values of $k$. The global probability is maximised at $24.7$\% for $\alpha=-6.5$, $k=1$ (Poisson), $\gamma=-1.4$, and $R(F>1\,{\rm Jy}\,{\rm ms})=4.27 \cdot 10^{-2}$, although there is a broad maximum for small values of $\gamma$, $\alpha$, and large values of $k$. Both the steep spectrum and the Poissonian burst time distribution is consistent with the analysis of \citet{Kumar2019}.

At the nominal value of $k=0.34$, the probabilities vary little in $\alpha$--$\gamma$ space, with only large values of $\gamma$ strongly disfavoured. 

Interestingly, for $k \le 0.2$, less-negative values of $\alpha$ become favoured. This is because it becomes more plausible to have a higher expected detection rate for the $>1$\,GHz Parkes and GBT observations, which just happen to miss periods of outburst, and a lower expectation for GBT 800\,MHz observations, which luckily happens to barely catch two outbursts. Since ASKAP observations were widely spread in time, it becomes difficult to avoid these with unlucky outburst timing, so that $\gamma=-1.5$ becomes strongly favoured.

The estimated probabilities $p$ correspond to the likelihood of an observation given a particular hypothesis on $R$, $\alpha$, $\gamma$, and $k$. Confidence intervals can therefore be calculated using Wilks' theorem \citep{Wilks62}, which states that in the large-sample limit, the test statistic
\begin{eqnarray}
\Delta D \equiv 2 \left[\log p(R,\alpha,\gamma,k)_{\rm max} - \log p(\alpha,R,\gamma,k)_{\rm true} \right] \sim \chi^2_4, \label{eq:D}
\end{eqnarray}
where the subscript `max' indicates the parameter values maximising the probability $p$, and `true' indicates the true value of these parameters. The $R$--$\alpha$--$\gamma$--$k$ space sets the degrees of freedom for the $\chi^2$ distribution to four. This is used to generate the confidence regions in Fig.~\ref{fig:caseC}. In Appendix~\ref{sec:chi2}, we show that this procedure results in slightly conservative limits.

Marginalising over all other parameters, the allowed ranges at 68\% confidence are $k>0.4$, $\alpha < -1$, with no constraints on $\gamma$ ($\gamma=-0.5$ is barely allowed for $\alpha=-8.0$, $k=1$). At 90\% C.L., only $k=0.1$ can be excluded. 

The nominal parameter set of $k=0.34$, $\gamma=-0.9$, $\alpha=-1.5$ lies on the 90\% exclusion level. The rate maximising this probability is $R(F>1\,{\rm Jy\,ms})=0.9\,{\rm day}^{-1}$, significantly less than that of FRB~121102.

\section{Discussion}
\label{sec:discussion}

\subsection{Absolute rates}
\label{sec:absolute_rates}

\begin{table}
    \centering
    \begin{tabular}{l|c| ccc}
    && \multicolumn{3}{|c}{$R_{90} (E>10^{39}\,{\rm erg})$ [day$^{-1}]$} \\
    FRB & $z_{\rm max}$ & $\gamma=-0.7$ & $\gamma=-0.9$ & $\gamma=-1.1$ \\
    \hline
170107 & 0.517 & 1.5 & 2.3 & 2.8 \\
170416 & 0.441 & 4.3 & 8.8 & 14 \\
170428 & 0.855 & 7.1 & 18 & 39 \\
170707 & 0.179 & 1.7 & 3.3 & 4.7 \\
170712 & 0.251 & 4.2 & 5.1 & 7.2 \\
170906 & 0.322 & 0.956 & 1.7 & 2.4 \\
171003 & 0.387 & 1.5 & 2.0 & 2.4 \\
171004 & 0.243 & 0.42 & 0.499 & 0.515 \\
171019$^\dag$ & 0.385 & $1_{-0.7}^{+1.3}$ & $1.4_{-0.9}^{+1.7}$ & $1.5_{-0.9}^{+2.0}$ \\
171020 & 0.0636 & 0.181 & 0.12 & 0.0571 \\
171116 & 0.525 & 2.4 & 6.2 & 11 \\
171213 & 0.107 & 0.248 & 0.245 & 0.154 \\
171216 & 0.149 & 1.9 & 3.5 & 5.2 \\
180110 & 0.611 & 4.7 & 12 & 19 \\
180119 & 0.333 & 1.0 & 1.8 & 2.6 \\
180128.0 & 0.368 & 1.5 & 2.2 & 2.5 \\
180128.2 & 0.417 & 6.5 & 16 & 36 \\
180130 & 0.279 & 0.824 & 1.7 & 2.0 \\
180131 & 0.56 & 4.2 & 9.6 & 17 \\
180212 & 0.115 & 0.228 & 0.188 & 0.117 \\
180315 & 0.401 & 6.5 & 6.7 & 6.3 \\
180324 & 0.358 & 5.3 & 5.4 & 4.5 \\
180417 & 0.398 & 33 & 40 & 36 \\
180430 & 0.206 & 1.5 & 0.906 & 0.49 \\
180515 & 0.29 & 7.9 & 7.6 & 6.2 \\
180525 & 0.32 & 3.9 & 3.6 & 3.4 \\
180924$^*$ & 0.3214 & 3.2 & 2.8 & 2.2 \\
    \end{tabular}
    \caption{Estimated maximum redshifts, $z_{\rm max}$, of each FRB, and corresponding 90\% C.L.\ upper limits on the intrinsic FRB rate $R_0$ (also given in Fig.\,\ref{fig:absolute_lim}) as a function of $\gamma$, for $k=0.34$ and other standard parameters. $^\dag$Best-fit repetition rate, and 90\% confidence limits. $^*$As it has been localised, the true redshift of FRB~180924 is used.}
    \label{tab:absolute_results}
\end{table}

A distance estimate to each FRB is required in order to translate our limits on rates above a given fluence observed at Earth into limits on the intrinsic rate above some energy. However, only one FRB in the sample, FRB~180924, has a confidently identified host \citep[at $z=0.3214$;  ][]{2019Sci...365..565B}. Nonetheless, a maximum distance to each can be estimated by attributing all non-Milky Way DM contributions to the intergalactic medium (IGM). Using the NE2001 model of \citet{CordesLazio01}, attributing a halo contribution equal to the minimum of 50\pccc\ \citep{ProchaskaZheng2019}, and ignoring any host galaxy contribution, allows the DM--$z$ relation due to the IGM from \citet{Inoue2004} to be used to estimate a maximum redshift, $z_{\rm max}$. These are given in Table~\ref{tab:absolute_results}. Note that limits on the intrinsic FRB behaviour become weaker as the assumed distance to the source increases, so that using $z=z_{\rm max}$ leads to upper limits on the intrinsic rate.

\begin{figure}
    \centering
    \includegraphics[width=\columnwidth]{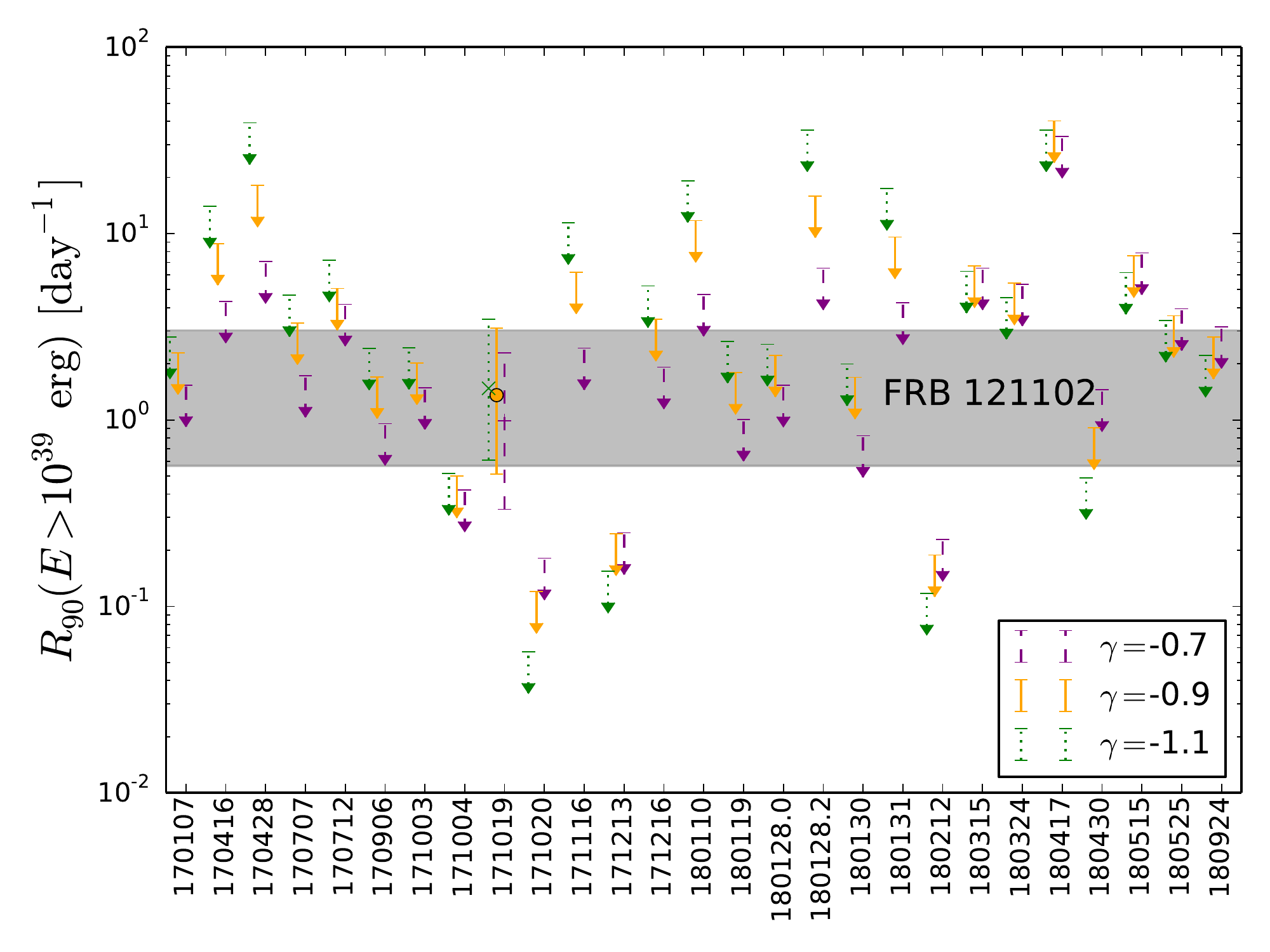}
    \caption{Points: upper limits at 90\% confidence on the intrinsic rate $R$ above an energy threshold of $E=10^{38}$\,erg, for three different values of $\gamma$. This is compared to the range of values estimates for FRB~121102 (grey shaded region --- see Section~\ref{sec:model}).}
    \label{fig:absolute_lim}
\end{figure}

The observational limits in Table~\ref{tab:upper_limits} are rates above 1\,{\rm Jy\,ms}, which translates to an energy threshold $E$ given by the fluence-energy relation of \citet{Macquart2018b},
\begin{eqnarray}
E & = & 4 \pi D_L^2 F \frac{\Delta \nu}{(1+z)^{2+\alpha}},
\label{eq:Fnu}
\end{eqnarray}
for luminosity distance $D_L$, assumed bandwidth $\Delta \nu=336$\,MHz, and a burst occupying the entirety of $\Delta \nu$ with spectral dependence $\alpha$. The intrinsic rate $R$ is also increased by $1+z$ to account for time dilation. The final limits are scaled to a rate above $E=10^{39}$\,erg using $\gamma=-0.9 \pm 0.2$. The results are shown in Fig.~\ref{fig:absolute_lim}.

For FRBs with high DM/$z_{\rm max}$, the limiting fluence of 1\,Jy\,ms translates to energies $>10^{39}$\,erg, so that limiting rates scaled to this threshold then get weaker as $\gamma$ decreases. The converse is true for low-DM FRBs. The largest  effect of $\gamma$ is therefore a factor of $5.5$ difference in $R_{90}$ for the highest-DM FRB in the sample, 170428, with a DM of 991.7\pccc. This effect dominates over variation in limits due to $0.29 \le k \le 0.4$. For very bursty distributions ($k=0.1$), the limits become significantly stronger when observations are clustered about the initial detection, and weaker when they are not. However, such behaviour is at odds with the behaviour of the three best-studied repeaters, FRB~121102 \citep{Oppermannetal2018}, FRB~180814.J0422+73 \citep{CHIME2019b}, and FRB~180916.J0158+65 \citep{2020arXiv200110275T}, so from hereon we quote only limits for $k=0.34$, equivalent to $0.29 \le k \le 0.4$.

The range of rates measured for FRB~121102 is shown as a grey band spanning 4.5--24\,day$^{-1}$ above $10^{38}$\,erg, covering estimates from \citet{Law2017}, \citet{Gajjaretal2018}, \citet{Oppermannetal2018}, and \citet{James2019b}, scaled to rates above $10^{39}$\,erg using $\gamma=-0.9$.
A total of 4 FRBs in our sample can be excluded at 90\% confidence as being repeating FRBs with repetition rates similar to that of FRB~121102. For a further seven, part of the rate range of FRB~121102 can be excluded.

For FRB~171019, its maximum redshift, $z_{\rm max}=0.385$, is much greater than that of FRB~121102, with $z = 0.193$  \citep{Tendulkar2017}. Hence, despite its observed rate being lower, its intrinsic rate may be identical to that of FRB~121102. It may, however, be much closer, and hence have a significantly lower intrinsic rate.

\subsection{Do all Fast Radio Bursts repeat?}

Other works have examined the question of whether or not all FRBs repeat similarly to FRB~121102. \citet{2018ApJ...854L..12P} consider individual bursts detected by Parkes, and compare the limits these once-off bursts set on the wait time between bursts $\Delta t$, and the flux ratio between successive bursts $S_{i}/S_{i+1}$, with the measured values for FRB~121102. The authors cite an absence of singly observed bursts in the space $S_{i}/S_{i+1}<1$, $\Delta t \lesssim 10^4$\,s as evidence for distinct populations. However, for a singly observed burst, the non-detection of a second ($i+1$) burst necessarily limits its flux $S_{i+1}$ to be less than that of the observed burst, $S_i$, \emph{by definition}. However, the non-observation of any \emph{preceding} burst also requires a point at $S_{i-1}/S_{i}<1$. The omission of points corresponding to preceding bursts creates a bias, resulting in an apparent, but illusory, disparity between singly observed bursts and those from FRB~121102. Furthermore, \citet{2018ApJ...854L..12P} do not account for the greater distance at which the Parkes sample of FRBs is detected \citep{Shannonetal2018}. Objects intrinsically identical to FRB~121102, but located at a greater distance, will exhibit a lower apparent rate, thus accounting for the absence of bursts in the interval $\Delta t \lesssim 10^4$\,s. By not accounting for distance effects, their result leaves open the possibility that Parkes FRBs come from objects intrinsically identical to FRB~121102, but which are generally more distant.

This is the conclusion of \citet{2016MNRAS.461L.122L}, who use a cosmological FRB source evolution proportional to the core-collapse supernova rate, and find that Parkes data are consistent with all repeaters being intrinsically similar to FRB~121102 at the 5--30\% level.

Our observations here exclude this possibility, since four FRBs in the CRAFT survey cannot repeat with the regularity of FRB~121102.

This does not necessarily mean that they do not repeat at all. Models of FRBs powered by young neutron stars \citep{cordesNS,Connor,Pen} or magnetar flares \citep{2010vaoa.conf..129P,Thornton,Kulkarni,2017ApJ...841...14M} would be expected to produce fewer, weaker bursts as they age through spin-down or magnetic field decay. This should then produce a population $\Phi$ of repeating FRBs with a distribution of repetition rates, similarly to the observed population of pulsars and magnetars. Indeed, in their specific model, \citet{2016MNRAS.461L.122L} find some evidence that the mean repetition rate of FRBs must be less than FRB~121102.

\citet{James2019b} argues that a rate distribution according to $\Phi(R) \sim R^{-2} dR$ is consistent with the number of single bursts observed in the CRAFT lat50 survey of \citet{Shannonetal2018}. Note this is not the distribution of rate as a function of energy for a given FRB, but the distribution of rates above a fixed energy over the population of FRBs.

In this model, FRB~121102 would be a rare, rapidly repeating object, as may be FRB~180814.J0422+73 \citep{CHIME2019b} and FRB~180916.J0158+65 \citep{CHIME2019c}, while the remaining seven repeating CHIME FRBs  may be more numerous, less-frequently repeating objects. Whether or not the model is quantitatively consistent with the observed number of both once-off and repeating CHIME FRBs would require more-detailed estimates of the CHIME survey's sensitivity, sky coverage, and effective observing time than are currently present in the literature.

Is this model consistent with the follow-up observations presented here? While the volumetric number density of repeating objects scales as $R^{-2} dR$, the number of bursts produced by each FRB scales with $R$ by definition. Hence, the probability that a burst comes from an FRB with intrinsic rate $R$ scales as $R^{-1} dR$. That is, each observed single burst has equal probability of being attributable to a given range in $d \log R$. To detect one repeating FRB, and exclude four, from having a repetition rate similar to that of FRB~121102, suggests that at most 60\% of all repeating FRBs would be expected to have this rate (90\% C.L.). Hence, the rate distribution for repeating FRBs must extend over at least two orders of magnitude in repetition rate, given the expected flatness in $\log (R)$. In other words, while all FRBs may indeed repeat --- it is, after all, observationally impossible to exclude an arbitrarily low repetition rate --- a sizeable fraction of FRBs must repeat with a low rate, or else come from a separate population of once-off progenitor events.

\subsection{Model dependence}

Our limits on the repetition rate $R$ have been calculated over a broad parameter space in burst spectrum ($\alpha$), energy dependence of the burst rate ($\gamma$), and time-clustering ($k$). Nonetheless, there is no guarantee that the true behaviour of repeating FRBs lies within this space. How robust are our results to deviations from our model?

Firstly, regardless of the validity of the Weibull distribution as a quantitative model for burst wait times, these observations show evidence against clustering. The primary evidence is viewing many one-off bursts, where clustering of any form would tend to favour either viewing many bursts, or none at all. The detection of repeat bursts from FRB~171019 over a broad spread of observation times also lends credence to this. 

Of particular note is that \citet{2020arXiv200110275T} have recently detected periodic emission from FRB~180916.J0158+65. Should an FRB observed by ASKAP behave similarly, it must necessarily have been observed in an active state. While we have not set limits as a function of potential periodicity between active and inactive states, any such limits on the time-averaged rate will be stronger than those presented here.

Secondly, these observations cover a relatively small spectral range, between $720$ and $1900$\,MHz, and we emphasise that all rates are quoted relative to ASKAP observation parameters at $1.3$\,GHz. This both makes our conclusions more robust to spectral dependencies in FRB behaviour, but also completely insensitive to effects outside this range. In particular, the power-law spectral model does not appear to extend down to $184$\,MHz \citep{Sokolowskietal2018}. Furthermore, for spectral models where most bursts are expected below 1\,GHz, the limited time-coverage of GBT 820\,MHz observations will make limits more sensitive to the time structure of bursts. This is not the case for bursts above 1\,GHz, where observations have a more uniform time-coverage.

Thirdly, while the energy dependence of the burst rate is consistent between FRB~121102 and the entire FRB population, it is clearly possible for a single FRB to exhibit properties that deviate significantly from the population mean. An example is the unusually steep spectral index for FRB~171019 considered here. Similarly, any given FRB could incur an excess DM and hence be located at a significantly lower redshift than is assumed when calculating absolute limits on the rate. Therefore, it is feasible that a small number of FRBs could violate the upper limits derived here. However, as a whole sample, the derived limits are relatively robust.

\section{Conclusions}

We have used the results of a survey of 27 ASKAP FRBs with the Robert C.\ Byrd Green Bank Telescope (GBT) and Parkes telescope to investigate FRB repetition. Only one FRB, 171019, has been detected to repeat, the details of which have already been reported by \citet{Kumar2019}. We have used a simulation of repeating FRBs, combined with exact observation parameters, to set limits on the repetition properties of these 27 objects. In particular, we allow for clustered distributions of burst arrival times.

For four of the 26 FRBs not observed to repeat, we can exclude repetition rates comparable to that of FRB~121102, i.e.\ $R(E>10^{39}\,{\rm erg}) < 0.5\,{\rm day}^{-1}$. This assumes burst fluence indices $-1.1 \le \gamma \le -0.7$ and arrival time clustering $0.29 \le k \le 0.4$, consistent with observations of known FRBs. For FRB~171019, the parameters of FRB~121102 estimated by \citet{Law2017} and \citet{Oppermannetal2018} are only consistent with observations at the $\sim10$\% level. Clustering of burst arrival times are disfavoured, but cannot be excluded.

Our results --- even including the one detected repeating object --- set strong limits on the model of all bursts being attributable to repeating FRBs, with at most 60\% (90\%\,C.L.) of these FRBs having an intrinsic burst distribution similar to FRB~121102.
We cannot exclude however that individual FRBs may repeat at much higher rates in parts of the spectrum unprobed by these observations, e.g.\ < 700\,MHz, or > 2\,GHz, or do so with burst energy distributions more complex than the power laws investigated here.

\section*{Acknowledgements}

The Parkes radio telescope is part of the Australia Telescope National Facility which is funded by the Australian Government for operation as a National Facility managed by CSIRO.
Part of this work was performed on the OzSTAR national facility at Swinburne University of Technology. OzSTAR is funded by Swinburne University of Technology and the National Collaborative Research Infrastructure Strategy (NCRIS).
The Australian SKA Pathfinder is part of the Australia Telescope National Facility which is managed by CSIRO. Operation of ASKAP is funded by the Australian Government with support from the National Collaborative Research Infrastructure Strategy. ASKAP uses the resources of the Pawsey Supercomputing Centre. Establishment of ASKAP, the Murchison Radio-astronomy Observatory and the Pawsey Supercomputing Centre are initiatives of the Australian Government, with support from the Government of Western Australia and the Science and Industry Endowment Fund. We acknowledge the Wajarri Yamatji people as the traditional owners of the Observatory site.
The Green Bank Observatory is a facility of the National Science Foundation operated under cooperative agreement by Associated Universities, Inc. Work at NRL is supported by NASA.
R.S.\ acknowledges support through ARC grants FL150100148 and CE170100004.
This research has made use of NASA's Astrophysics Data System Bibliographic Services. This research made use of Python libraries \textsc{Matplotlib} \citep{Matplotlib2007}, \textsc{NumPy} \citep{Numpy2011}, and \textsc{SciPy} \citep{SciPy2019}.



\bibliographystyle{mnras}
\bibliography{ASKAP_reps}

\appendix
\section{Distribution of the test-statistic $D$}
\label{sec:chi2}

In this work, we use Wilks' theorem \citep{Wilks62} to assume a $\chi^2$ distribution for the test-statistic $D$ given in equation~(\ref{eq:D}). For the analysis of Section~\ref{sec:FRB171019}, a likelihood maximisation is performed over the parameter set $R$, $\alpha$, $\gamma$, and $k$, such that $D \sim \chi^2_4$. Wilks' theorem states that $D$ will reach this asymptotic form only as the number of data points used in the maximisation tends to infinity. In this case, with only two bursts observed from FRB~171019, it is not at all clear that this asymptotic forms has been reached. Hence, we perform a toy simulation to test the validity of our assumption.

\begin{figure}
    \centering
    \includegraphics[width=\columnwidth]{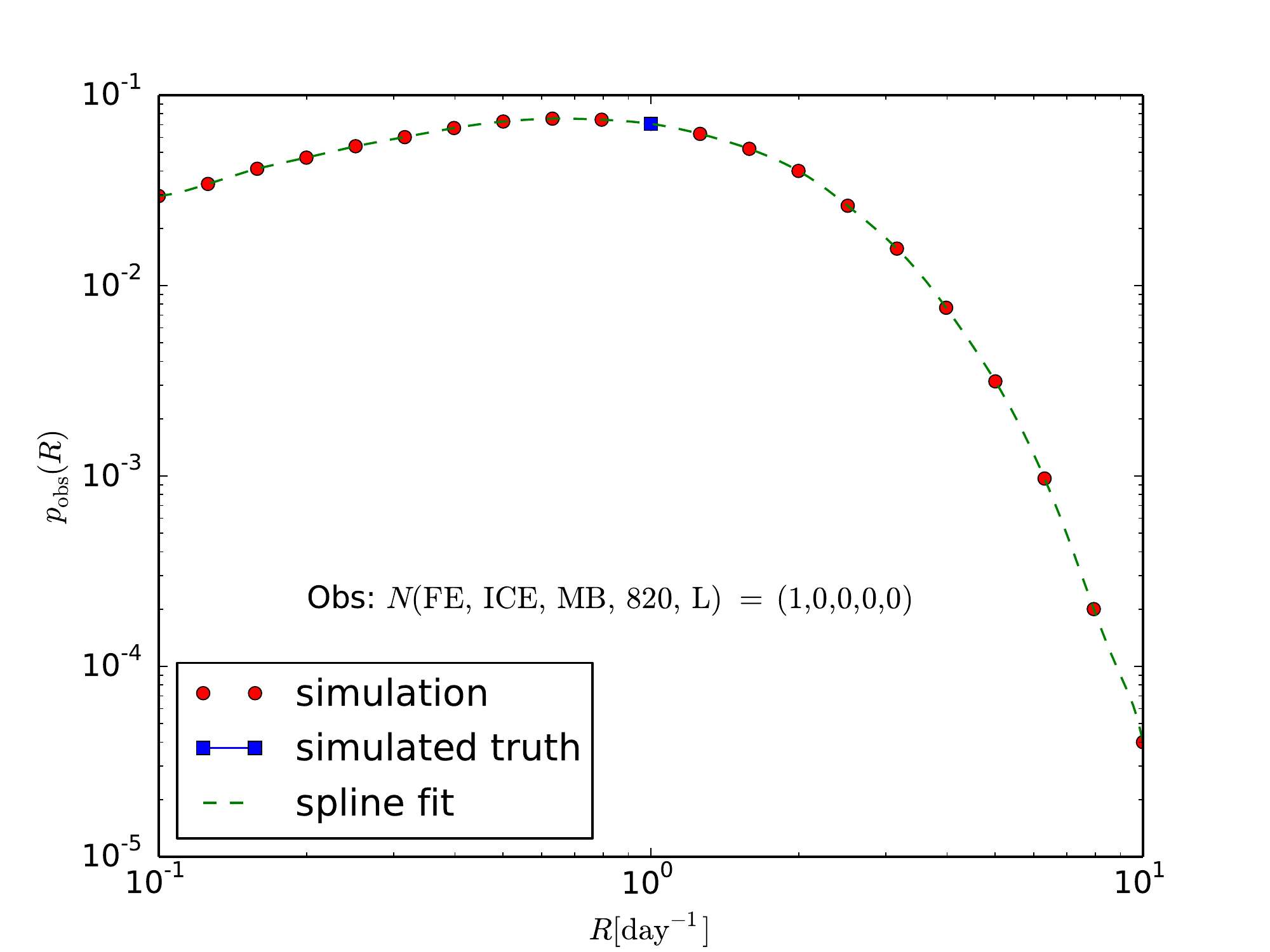}
    \caption{Red dots: simulated probability of observing one repeat burst from FRB~171019 with ASKAP Flye's Eye mode (N(FE)=1), and none with other instruments, as a function of repetition rate $R$, for $\gamma=-0.9$, $k=0.34$, $\alpha=-1.5$. The simulated true value of $R$ used to generate this observation is shown in blue. The green dashed line shows the spline fit.}
    \label{fig:chi2_fit}
\end{figure}

We use a simplified case, and consider only the $R$ dimension, with other parameters fixed at $k=0.34$, $\gamma=-0.9$, $\alpha=-1.5$. We assume a true rate $R=1$\,day$^{-1}$, and simulate over the range $0.1 \le R \le 10$\,day$^{-1}$ for FRB~171019 using the same simulation of Section~\ref{sec:simulation_method}. Only the total number of bursts observed by each of the five receivers in Table~\ref{tab:telescope_properties} is recorded, i.e.\ no timing information is assumed. For each simulated observation, $D$ is calculated by fitting cubic splines to the simulated probabilities of that result as a function of $\log R$. This allows the probability distribution $p({\rm obs}|R)$ to be smooth. Results where no repeat bursts were simulated were discarded. An example of the fitting procedure is shown in Fig.~\ref{fig:chi2_fit}, while the resulting distribution of $D$ is compared to a $\chi^2_1$ distribution in Fig.~\ref{fig:chi2}.

\begin{figure}
    \centering
    \includegraphics[width=\columnwidth]{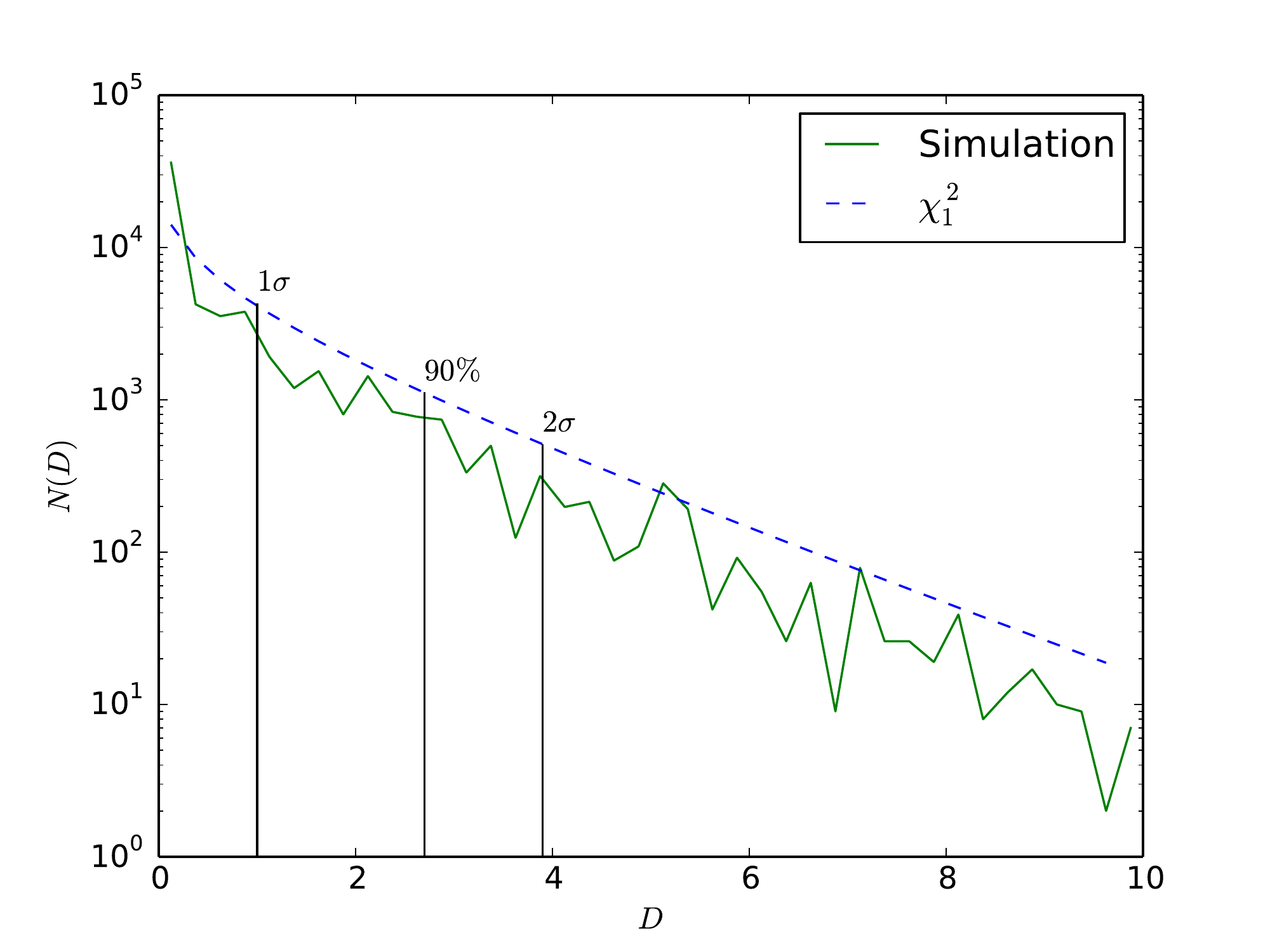}
    \caption{Simulated histogram of the test statistics $D$, compared to its expected $\chi^2_1$ distribution, scaled to the number of simulations. Also shown (black lines) are characteristic cut-off values associated with 68\% ($1\sigma$), 90\%, and 95\% ($2 \sigma$) confidence levels. }
    \label{fig:my_label}
\end{figure}

From Fig.~\ref{fig:chi2}, it is evident that the broad form of $D$ is very similar to that of a $\chi^2_1$ distribution. However, there is an excess near $D=0$, and a deficit at larger values. A possible cause is the quantisation of FRB observations, i.e.\ only integer numbers of FRBs can be observed. Importantly, the true distribution of $D$ lies at lower values than expected, meaning that confidence intervals set by assuming a $\chi^2_1$ distribution will suffer from over-coverage, e.g.\ a 90\% confidence limit may in fact be a 92\% C.L. Extrapolating to the multi-dimensional cases treated in this work, we expect true parameter values to lie outside our 90\% confidence regions less than 10\% of the time.

\bsp	
\label{lastpage}
\end{document}

%% file: address.tex
$^{1}$International Centre for Radio Astronomy Research, Curtin University, Bentley, WA 6102, Australia \\
$^{2}$Centre for Astrophysics and Supercomputing, Swinburne University of Technology, Mail H30, PO Box 218, VIC 3122, Australia \\
$^{3}$CSIRO Astronomy $\&$ Space Science, Australia Telescope National Facility, P.O. Box 76, Epping, NSW 1710, Australia \\
$^4$Space Science Division, Naval Research Laboratory, Washington, DC 20375, USA \\
$^{5}$Department of Physics and Astronomy, West Virginia University, P.O. Box 6315, Morgantown, WV 26506, USA \\
$^{6}$Center for Gravitational Waves and Cosmology, West Virginia University, Chestnut Ridge Research Building, Morgantown, WV 26505 \\
$^{7}$Dunlap Institute for Astronomy and Astrophysics, University of Toronto, 50 St. George Street, Toronto, ON M5S 3H4, Canada \\
$^8$Department of Astronomy, University of California Berkeley, Berkeley CA 94720, USA \\
$^9$Sydney Institute for Astronomy, School of Physics, University of Sydney, Sydney, NSW 2006, Australia \\
$^{10}$ARC Centre of Excellence for Gravitational Wave Discovery (OzGrav)



